\documentclass[pra,aps,showpacs,twocolumn,floatfix,superscriptaddress,showkeys]{revtex4}  
\usepackage{graphicx}
\usepackage[ansinew]{inputenc}
\usepackage{array}
\usepackage{color}
\usepackage{amsmath}
\usepackage{amsxtra}
\usepackage{amstext}
\usepackage{amssymb}
\usepackage{latexsym}
\usepackage{dsfont}
\usepackage{verbatim}
\usepackage{comment}
\usepackage{epstopdf}

\begin{document}

\title
{Engineering nonlinear response of nanomaterials using Fano resonances}

\author{Deniz Turkpence}
\affiliation{Institute of Nuclear Sciences, Hacettepe University, 06532, Ankara, Turkey}
%\affiliation{Center for Advanced Researches, K{\i}rklareli University, 39020 Karah{\i}d{\i}r, K{\i}rklareli, Turkey}
\author{Gursoy B. Akguc}
\affiliation{Department of Physics, Bilkent University, 06800, Ankara, Turkey}
\author{Alpan Bek}
\affiliation{Department of Physics, Middle East Technical University, 06800, Ankara, Turkey}
\affiliation{The Center for Solar Energy Research and Applications (GUNAM), Middle East Technical University, 06800, Ankara, Turkey }
\author{Mehmet Emre Tasgin}
\affiliation{Institute of Nuclear Sciences, Hacettepe University, 06532, Ankara, Turkey}
%\affiliation{Department of Physics, Bilkent University, 06800, Ankara, Turkey}
%\affiliation{to whom correspondence should be addressed.}

\date{\today}

\begin{abstract}
We show that, nonlinear optical processes of nanoparticles can be controlled by the presence of interactions with a molecule or a quantum dot. By choosing the appropriate level spacing for the quantum emitter, one can either suppress or enhance the nonlinear frequency conversion. We reveal the underlying mechanism for this effect, which is already observed in recent experiments: (i) Suppression occurs simply because transparency induced by Fano resonance does not allow an excitation at the converted frequency. (ii) Enhancement emerges since nonlinear process can be brought to resonance. Path interference effect cancels the nonresonant frequency terms. We demonstrate the underlying physics using a simplified model, and we show that the predictions of the model are in good agreement with the 3-dimensional boundary element method (MNPBEM toolbox) simulations. Here, we consider the second harmonic generation in a plasmonic converter as an example to demonstrate the control mechanism. The phenomenon is the semi-classical analog of nonlinearity  enhancement via electromagnetically induced transparency.
\end{abstract}



%42.50.Gy	Effects of atomic coherence on propagation, absorption, and amplification of light; electromagnetically induced transparency and absorption
%42.65.Ky	Frequency conversion; harmonic generation, including higher-order harmonic generation (see also 42.79.Nv Optical frequency converters)
%73.20.Mf	Collective excitations (including excitons, polarons, plasmons and other charge-density excitations)

\maketitle

%%%%%%%%%%%%%%%%%%%%%%%%%%%%%%%%%%%%%%%%%%%%%%%%%%%%%%%%%%%%%%%%%%%%%%%%%%%%%%%%%%%%%%%%%%%%%%%%%%%%%%%%%%%
%%%%%%%%%%%%%%%%%%%%%%%%%%%%%%%%%%%%%%%%%%%%%%%%%%%%%%%%%%%%%%%%%%%%%%%%%%%%%%%%%%%%%%%%%%%%%%%%%%%%%%%%%%%
%%%%%%%%%%%%%%%%%%%%%%%%%%%%%%%%%%%%%%%%%%%%%%%%%%%%%%%%%%%%%%%%%%%%%%%%%%%%%%%%%%%%%%%%%%%%%%%%%%%%%%%%%%%
%%%%%%%%%%%%%%%%%%%%%%%%%%%%%%%%%%%%%%%%%%%%%%%%%%%%%%%%%%%%%%%%%%%%%%%%%%%%%%%%%%%%%%%%%%%%%%%%%%%%%%%%%%%
%%%%%%%%%%%%%%%%%%%%%%%%%%%%%%%%%%%%%%%%%%%%%%%%%%%%%%%%%%%%%%%%%%%%%%%%%%%%%%%%%%%%%%%%%%%%%%%%%%%%%%%%%%%
%%%%%%%%%%%%%%%%%%%%%%%%%%%%%%%%%%%%%%%%%%%%%%%%%%%%%%%%%%%%%%%%%%%%%%%%%%%%%%%%%%%%%%%%%%%%%%%%%%%%%%%%%%%
\section{Introduction}

Resonant interaction of metal nanoparticles  (MNPs) with optical light provides a tool for the strong localization of electromagnetic field \cite{stockman-review}. Intensity enhancements as high as $10^5$ can be achieved \cite{stockman-review,QDinducedtrans} within the localized surface plasmon-polariton (PP) fields, in terms of coupled oscillations of surface electrons and the localized optical field \cite{nanooptics-book}. Such orders of magnitude increase in the intensity leads to the emergence of optical nonlinearities \cite{KauranenNature2012} , e.g. enhanced Raman scattering \cite{Sharma2012}, four wave mixing \cite{GenevetNanoLett2010} and second harmonic generation (SHG) \cite{MartinNanoLett2013,silencing_enhancementSHG,WunderlichOptExp2013,Gao-AccChemRes-2011,WalshNanoLett2013,SinghNanotech2013}. 

Emergence of nonlinear processes can be both desirable or unwanted depending on the operating properties of the fabricated device. As an example, plasmon-polariton mediated surface enhancement is successfully used to achieve Raman imaging of materials \cite{KneippPRL1997,AngelesNatureNano2010}. The nonlinear response of the media can also be utilized for optical switching \cite{switch}. The SHG process can enhance the absorption efficiency in photovoltaic devices \cite{PV}, may increase the coherence time (length) of the field \cite{BECsqueezed1,BECsqueezed2,metasginNanoscale2013} as well as being able to generate entangled photon pairs \cite{plasmon-entanglement}.  

Despite such advantages, nonlinear conversion may be undesirable in other devices. Raman scattering process in fiber-optic cables causes loses in the signal and limits the number of channels that could be used for a given bandwidth \cite{avoidNL1,avoidNL2,avoidNL3,avoidNL4}. Similarly, nonlinear effects can decrease the quality factor of microwave cavities \cite{hqcavity3,hqcavity4}. In addition, one may require the operation of a device in the linear regime even for higher input powers. Because nonlinearities may cause unexpected chaotic behavior for the long term operation \cite{chaosbook}.

Besides the emergence of nonlinearities, Fano resonances --analogous to electromagnetically induced transparency \cite{Scullybook}-- has also been observed in plasmonic excitations of MNPs \cite{QDinducedtrans,QCfano1,QCfano2,QCfano3,QCfano4,QCfano5,QCfano6,QCfano7,QCfano8,QCfano9}. The attachment of a quantum oscillator [e.g. a molecule or a quantum dot (QD)] to a MNP strongly modifies the optical response of the hybrid material \cite{metasginNanoscale2013,QDinducedtrans,spaser,Pfeiffer2010,Zhao2011,Anger2006}. The presence of a quantum oscillator with small decay rate induces a weak hybridization. Hybridization is weak in the sense that frequency splitting in the MNP resonance is very small compared to the spectral width of the resonance. This introduces two possible excitation paths for the absorption/polarization of the incident light, which are unresolvable.  Both excitation frequencies lie within the frequency window of MNP resonance and interfere destructively \cite{Alzar,Soukoulis2012}. There emerges a transparency window centered about $\omega_{eg}$, where polarization of MNP-quantum oscillator hybrid system is avoided \cite{metasginNanoscale2013}. Such resonances are observed as long as the decay/damping rate of one of the oscillators is significantly small compared to the second one \cite{metasginNanoscale2013,Alzar,Soukoulis2012}.
 
\begin{figure}% [htb!]
\includegraphics[width=3.4in]{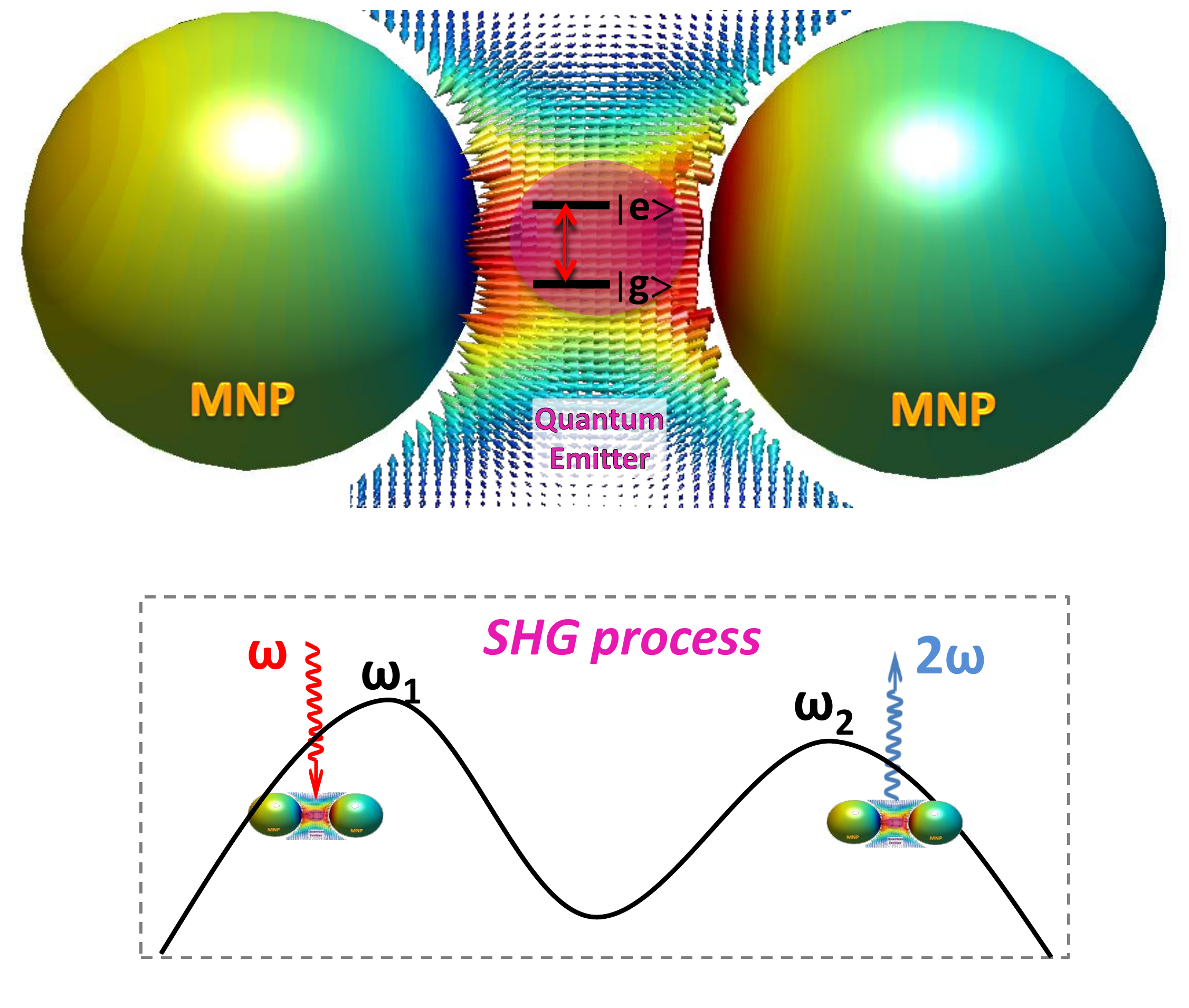}
\caption{{\it Top:} A quantum emitter (purple) with a small decay rate is placed at the center of a MNP dimer \cite{QCfano1}. The polarization of the plasmon-polariton (PP) modes strongly localizes the incident field to the center (see the field vectors). Field enhancement gives rise nonlinear processes, e.g. second harmonic generation (SHG). The two MNPs are chosen in the same size only for the purposes of demonstration. {\it Bottom:} The incident planewave field ($\epsilon_p e^{-i\omega t}$) drives the $\hat{a}_1$ PP polarization mode (resonance $\omega_1$) of the dimer. The intense localized polarization field of $\hat{a}_1$, oscillating with $\omega$, gives rise to SHG \cite{2plas1phot}. This process induces oscillations ($e^{-i2\omega t}$) in the second PP mode $\hat{a}_2$ whose resonance is $\omega_2$. The quantum oscillator (level spacing $\omega_{eg}\simeq 2\omega$) interacts with the field of the $\hat{a}_2$ polarization mode. Quantum oscillator (emitter) is chosen to have no SH response \cite{EYFP} to $\omega$. The observation of the SH light occurs due to the radiative decay of the $\hat{a}_2$ PP mode \cite{BouhelierPRL2005,2plas1phot,BeversluisPRB2003,PohlScience2005}.}
\label{fig1}
\end{figure}

In this paper, we show that it is possible to manage the nonlinear behavior of a material using the path interference effects. As an example, we consider the following system.  We place a quantum oscillator (QD, molecule or a nitrogen vacancy center) at the hot-spot of a MNP dimer (see Fig.~\ref{fig1}, {\it top}) which has a low decay rate ($\gamma_{eg}$) compared to the MNP \cite{QCfano1} ($\gamma_{1,2}$). The dimer has localized surface  plasmon-polariton (PP) resonances $\omega_1$ and $\omega_2$ (see Fig.\ref{fig1}, {\it bottom}) for the polarization field. The drive frequency $\omega$ and the second harmonic (SH) frequency $2\omega$ fall into the excitation range of the $\omega_1$ and $\omega_2$ polarization modes, respectively. Without the presence of the quantum oscillator, resonance of the SHG process occurs when $\omega_1=\omega$ and $\omega_2=2\omega$ (see the discussion in Sec.~\ref{sec:suppression}). We show that, (i) even in the resonance condition for SH conversion (that is $\omega_1=\omega$, $\omega_2=2\omega$), the presence of coupling to the quantum oscillator (emitter)  can suppress the nonlinear process several orders of magnitude. The factor of achievable suppression is inversely proportional to the square of the quantum decay rate and increases with the strength of the MNP-quantum oscillator coupling. A suppression factor of $\sim10^{-9}$ is possible when a high-quality (small decay rate) quantum oscillator (a QD), with spectral width of $10^9$ Hz,  is coupled to the MNP  (see Fig.~\ref{fig2}). This effect is observed, because cancellation of the two excitation paths does not allow polarization in the $\omega_2$ PP mode of the MNP dimer. Suppression is maximum when quantum level spacing is resonant to conversion frequency, $\omega_{eg}=2\omega$. (ii) On the other hand, a similar path cancellation effect can be adopted to kill the nonresonant term [($\omega_2-2\omega$)] that emerges when $\omega_2$ is not resonant to the SHG frequency $2\omega$. Without adjusting the resonances \cite{NordlanderNanoLett2004} of the dimer ($\omega_1$, $\omega_2$), the SHG process can be carried closer to resonance (Fig.~\ref{fig4}). These two effects together, enables the control over induction of the nonlinearities without the need for managing the properties of the material.

The effect of Fano resonances on the nonlinear conversion processes has already been studied both theoretically (using finite element simulations) \cite{BrevetnonlinearFanoPRB2012} and experimentally \cite{MartinNanoLett2013,silencing_enhancementSHG,WalshNanoLett2013}. However, an explicit demonstration [i.e. as in Eq.~(\ref{alph2}) below] of how path interference can tune the conversion has not been examined, yet. In an exciting recent work \cite{SinghNanotech2013}, a theoretical model of enhancement of the nonlinear response originating from a quantum dot by help of plasmonic particles was studied.  Our work, in contrast, provides a theoretical model of how the nonlinear response originating from a plasmonic (classical) system is enhanced by a quantum oscillator.

Emergence of the suppression phenomenon necessitates presence of coupling to a quantum oscillator with a small decay rate. However, enhancement effect may also be observed for coupled plasmonic resonators with broad absorption (emission) bands \cite{MartinNanoLett2013,BrevetnonlinearFanoPRB2012}.
  
We verify the emergence of such an enhancement by 3-dimensional (3D) boundary element method simulations using the MNPBEM toolbox \cite{MNPBEM} in Matlab (see Fig.~\ref{fig5}). Our simulations take the retardation effect into account. The simple model can predict the both the emergence and the spectral position of the SHG enhancement successfully. 
  
In a separate experiment of our research team \cite{BekSHG}, we observe the SH radiation from the hybrid system of composite MNPs which are decorated with dye molecules. SHG can originate only from the MNPs since the EYFP molecule \cite{EYFP} does not have a SH response to the drive frequency \cite{PS5}. Our simple model can easily predict [using Eq.s~(\ref{eq:timea})-(\ref{eq:timed})] an enhancement factor of $\sim$1000 in the experiment. Taking this enhancement factor into account, SH signal --from MNP clusters illuminated with a CW laser (40MW/$\text{cm}^2$ at the sample)-- reaches the values attained in the typical experiments \cite{bautistaNanoLett2012} where samples are illuminated with high peak intensity (60GW/$\text{cm}^2$) ultra-short lasers. In fact, the observed effect is the classical (or semi-classical) analog of nonlinear response enhancement obtained via electromagnetically induced transparency (EIT)-like atomic coherence \cite{imamogluEITnonlin}. In difference, the present system does not necessitate a microwave drive, which makes it efficient considering energy consumption.

Here, we present the method for the second harmonic generation process in a MNP dimer. However, method can be generalized to other nonlinear frequency generation processes as long as the two modes of the material can be resolved (see Sec.~\ref{sec:otherNL}). It is also possible to use the excitation modes of other nanoscale resonators \cite{hqcavity1,hqcavity2}.

The paper is organized as follows. In Sec.~\ref{sec:hamiltonian}, we describe the SHG process in the coupled system of a MNP dimer and a quantum oscillator. We introduce the Hamiltonian for the hybrid system. Nonlinear frequency conversion process is included in the second quantized Hamiltonian.  We derive the equations of motion for the system using the density matrix formalism for the quantum oscillator. We include the damping and quantum decay rates and the source driving the MNP dimer. In Sec.~\ref{sec:suppression}, we demonstrate that conversion process is suppressed for $\omega_{eg}\simeq2\omega$. In Sec.~\ref{sec:enhancement}, we present a contrary effect. The cancellation of the nonresonant terms leads to enhanced production of the SHG for the choice of $\omega_{eg}\simeq 1.98\omega$.  In Sec.~\ref{sec:MNPBEM}, we compare the results of our model with the 3D simulations which are based on the exact solutions of the Maxwell equations. In Sec.~\ref{sec:otherNL}, we discuss how the model can be adopted to other nonlinear processes. In Sec.~\ref{sec:classification}, we provide a classification for the types of Fano resonances and their relevance with EIT \cite{Scullybook}, for the sake of generalization of the enhancement phenomenon to other composite systems.   Sec.~\ref{sec:conclusion} includes our conclusions.

%%%%%%%%%%%%%%%%%%%%%%%%%%%%%%%%%%%%%%%%%%%%%%%%%%%%%%%%%%%%%%%%%%%%%%%%%%%%%%%%%%%%%%%%%%%%%%%%%%%%%%%%%%
%%%%%%%%%%%%%%%%%%%%%%%%%%%%%%%%%%%%%%%%%%%%%%%%%%%%%%%%%%%%%%%%%%%%%%%%%%%%%%%%%%%%%%%%%%%%%%%%%%%%%%%%%%
%%%%%%%%%%%%%%%%%%%%%%%%%%%%%%%%%%%%%%%%%%%%%%%%%%%%%%%%%%%%%%%%%%%%%%%%%%%%%%%%%%%%%%%%%%%%%%%%%%%%%%%%%%
%%%%%%%%%%%%%%%%%%%%%%%%%%%%%%%%%%%%%%%%%%%%%%%%%%%%%%%%%%%%%%%%%%%%%%%%%%%%%%%%%%%%%%%%%%%%%%%%%%%%%%%%%%%
%%%%%%%%%%%%%%%%%%%%%%%%%%%%%%%%%%%%%%%%%%%%%%%%%%%%%%%%%%%%%%%%%%%%%%%%%%%%%%%%%%%%%%%%%%%%%%%%%%%%%%%%%%%
%%%%%%%%%%%%%%%%%%%%%%%%%%%%%%%%%%%%%%%%%%%%%%%%%%%%%%%%%%%%%%%%%%%%%%%%%%%%%%%%%%%%%%%%%%%%%%%%%%%%%%%%%%%

\section{Modification of the nonlinear response}

In this section, we describe the response of a coupled MNP dimer-quantum oscillator system to a driving electromagnetic field. We shortly mention about the nature of couplings in the hybrid system and the mechanism for SHG on the MNP-dimer resonator. 

We give the effective Hamiltonian for the system and drive the equations of motion for the fields of the plasmon-polariton modes together with the excitation of the quantum oscillator. We find the equations governing the steady state values of the excitations to obtain the linear behavior of the hybrid system. Using these equations, we demonstrate the principle behind gaining control over the process of nonlinear frequency generation. We show that by choosing the appropriate level spacing ($\omega_{eg}$) for the quantum oscillator, one can either suppress and or enhance the non-linear frequency generation.

%%%%%%%%%%%%%%%%%%%%%%%%%%%%%%%%%%%%%%%%%%%%%%%%%%%%%%%%%%%%%%%%%%%%%%%%%%%%%%%%%%%%%%%%%%%%%%%%%%%%%%%%%%%
\subsection{Hamiltonian and equations of motion} \label{sec:hamiltonian}

We consider a system where a quantum oscillator (e.g. quantum dot \cite{QDattached}, molecule \cite{molattached} or a nitrogen-vacancy center \cite{Zhao2011,Anger2006}) is placed in the center of the MNP dimer. The two MNPs can still be coupled to each other, and be said to have dimerized, due to the small dimensions of the quantum oscillator \cite{QCfano1}. The two resonances of the MNP dimer $\omega_1$ and $\omega_2$ are relevant to the incident ($\omega$) and SH ($2\omega$) frequencies, respectively (see Fig.~\ref{fig1}, {\it bottom}). The resonance frequency of the $\hat{a}_2$ PP mode ($\omega_2$) is about the SH frequency $2\omega$, but not necessarily resonant with it.

The incident light, in the planewave mode with frequency $\omega$, couples strongly  to the $\hat{a}_1$ plasmon-polariton mode of the dimer. The direct coupling of light to the quantum oscillator is of negligible strength compared to the plasmon. Quantum oscillator couples to the localized plasmon-polariton field of the dimer. The hot-spots for the both plasmon-polariton modes emerge in the middle of the two MNPs \cite{HalasChemRev2011}, where the quantum oscillator is tightly placed. 

The dynamics of the total system is as follows. The incident planewave mode field ($\epsilon_p e^{-i\omega t}$) drives the first dimer mode $\hat{a}_1$ (resonance $\omega_1$) at the oscillation frequency $\omega$. The polarization of the plasmon-polariton (PP) excitation yields a localized strong electromagnetic field mode ($\hat{a}_1$) between the two MNPs. Such an enhancement  in the field gives rise to the emergence of nonlinear effect (e.g. SHG) in the electron gas \cite{2plas1phot,CiraciPRB2012,FinazziPRB2012,QCfano1}. Explicitly; the field (oscillating at $\omega$) trapped in the $\hat{a}_1$ PP polarization gives rise to second harmonic polarization oscillations ($2\omega$) \cite{2plas1phot,CiraciPRB2012,FinazziPRB2012,QCfano1,DykmanPRE1996,PanasyukPRL2008,HubnerPRB1994,KielichJMOpt1987,BrevetPRB2010,Georges-JOptSocAmB-2011,PanasyukPRB2011,Trugler-thesis-2011}  in the $\hat{a}_2$ PP polarization mode of the dimer. The quantum oscillator, whose level spacing is compatible with the SH oscillation frequency $\omega_{eg}\simeq 2\omega$, interacts with the polarization field of the $\hat{a}_2$ PP mode. The resonance of $\hat{a}_2$ mode is $\omega_2$. The field localization at the hot-spot provides strong interaction with the quantum oscillator. The SH light is observed through the radiative decay of the $\hat{a}_2$ PP mode \cite{BouhelierPRL2005,2plas1phot,BeversluisPRB2003,PohlScience2005}. We assume that the molecule does not have a SHG response to the drive frequency $\omega$. Such molecules exists; for example, see Fig. 2 in Ref.~\cite{EYFP}.

%Electromagnetic field is intensively localized at the hot-spot in the polarization of plasmon-polariton excitation. Such a field enhancement give rise to both the emergence of nonlinear effect (SHG) and provides strong interaction with the quantum object. Explicitly, field trapped in the plasmon-polariton modes $(\hat{a}_1^\prime)^{\omega_1}$ gives rise to second harmonic plasmon-polariton mode $(\hat{a}_2^\prime)^{\omega_2}$. The SH light is observed through the radiative decay of this excitation mode ($\hat{a_2}$) \cite{.}. We assume that the molecule doesn't have a SH response to the drive frequency $(\omega)$. 

Here, we consider a simplified model for the hybrid system. We mainly aim to demonstrate the principles behind the control mechanism. In more realistic calculations \citep{QCfano4,QCfano5}, one has to consider complicating effects, such as the influence of the dielectric environment and exact spatial distribution of the fields. However, the oscillators model \cite{metasginNanoscale2013,QCfano4} predicts the basic behavior of the MNPs combined with the quantum oscillators \cite{Zhao2011,Anger2006,spaser}.

The total Hamiltonian $(\hat{H})$ for the described system can be written as the sum of the energy of the quantum oscillator $(\hat{H_0})$, energy of the plasmon-polariton oscillations $(\hat{a_1}\:\text{,}\:\hat{a_2})$ of the MNP dimer $(H_d)$, the interaction of the quantum oscillator with the plasmon-polariton modes \cite{metasginNanoscale2013,QCfano1} $(H_{int})$ 
\begin{eqnarray}
\hat{H}_0=\hbar\omega_{\rm e} 
| {\rm e} \rangle \langle {\rm e} | + \hbar\omega_{\rm g}  | {\rm g} \rangle \langle {\rm g}| , \label{eq:H0} 
\\
\hat{H}_d=\hbar\omega_1\hat{a}_1^\dagger\hat{a_1}+\hbar\omega_2\hat{a}_2^\dagger\hat{a_2} , \label{eq:Hd}
\\
\hat{H}_{int}=\hbar(f_1\hat{a}_1^\dagger| {\rm g} \rangle \langle {\rm e} |+f_1^*\hat{a}_1| {\rm e} \rangle \langle {\rm g} |) \nonumber \\  
+\hbar(f_2\hat{a}_2^\dagger| {\rm g} \rangle \langle {\rm e} |+f_2^*\hat{a}_2| {\rm e} \rangle \langle {\rm g} |), \label{eq:Hint}
\end{eqnarray}
as well as the energy transferred by the pump source $(\omega)$, $\hat{H}_p$ and the second harmonic generation process among the plasmon-polariton fields $(\hat{H}_{sh})$ 
\begin{eqnarray}
\hat{H}_p=i\hbar(\hat{a}_1^\dagger\epsilon_p e^{-i\omega t} -\hat{a}_1\epsilon_p^* e^{i\omega t}) , \label{eq:Hp}
\\
\hat{H}_{sh}=\hbar\chi^{(2)}(\hat{a}_2^\dagger\hat{a}_1\hat{a}_1+\hat{a}_1^\dagger\hat{a}_1^\dagger\hat{a}_2) , \label{eq:Hsh}
\end{eqnarray}
respectively \cite{Scullybook,mandelwolf}. In Eq.~(\ref{eq:H0}), $\hbar\omega_e$ ($\hbar\omega_g$) is the excited (ground) state energy of the quantum oscillator. States ($| {\rm e} \rangle$), $| {\rm g} \rangle$ correspond to the (excited) ground levels of the quantum oscillator. $\hat{a}_1$, $\hat{a}_2$ are the plasmon-polariton excitations induced on the MNP dimer and $\hbar\omega_1$, $\hbar\omega_2$ are the corresponding energies for the oscillation modes. $f_1$ ($f_2$) is the coupling matrix element between the field induced by the $\hat{a}_1$ ($\hat{a}_2$) polarization mode of the MNP dimer and the quantum oscillator. Eq.~(\ref{eq:Hp}) describes the interaction of the light source (oscillates as $e^{-i\omega t}$) driving the plasmon-polariton mode with smaller resonance frequency $\omega_1$. In Eq.~(\ref{eq:Hsh}), the fields of two excitations in the low-energy plasmon-polariton mode ($\hat{a}_1$) combine to generate the field of a high energy plasmon-polariton mode. Stronger the second harmonic generated plasmon-polariton oscillations, the higher the number of emitted SHG photons $(2\omega)$. Because, $\hat{a}_2$ mode radiatively decay to $2\omega$ photon mode \cite{BouhelierPRL2005,2plas1phot}. Energy is conserved in the input-output process. The parameter $\chi^{(2)}$, in units of frequency, is proportional to the second harmonic susceptibility of the MNP dimer. 

We note that, one could also treat the SHG process as originating directly from the incident field, e.g. $\hat{H}_{sh}\sim ( \hat{a}_2^{\dagger} \epsilon_p^2e^{-i2\omega t}+c.c. )$. Even though the following results would remain unaffected, physically such a model would be inappropriate. Because, enhanced nonlinear processes emerge due to the electromagnetic field of the localized intense surface plasmon-polariton (polarization) mode \cite{BouhelierPRL2005,2plas1phot}. However, the mode of the incident field ($\omega$) is planewave.

We use the commutation relations (e.g. $i\hbar\dot{\hat{a}}=[\hat{a}\text{,}\hat{H}]$) in driving the equations of motions. After obtaining the dynamics in the quantum approach, we carry $\hat{a}_1\text{,}\;\hat{a}_2$ to   classical expectation values $\hat{a}_1\rightarrow\alpha_1\text{,}\;\hat{a}_2\rightarrow\alpha_2$. We introduce the decay rates for plasmon-polariton fields $\alpha_1$, $\alpha_2$. Quantum oscillator is treated within the density matrix approach.  Since we restrict ourselves to the classical properties of the fields, e.g. we do not not take squeezing into account, we could alternatively take the plasmon fields to be of classical nature in Eq.s~(\ref{eq:H0})--(\ref{eq:Hsh}). We could derive the equations of motion (\ref{eq:timea})-(\ref{eq:timeb}) by functional minimization. However, we choose to keep the operator notations for $\hat{a}_1$ and $\hat{a}_2$ quanta up to a certain derivation step in order to avoid incomplete modeling of the equations of motion.

The equations of motion take the form 
\begin{subequations}
\begin{align}
\dot{\alpha}_1=(-i\omega_1-\gamma_1)\alpha_1-i2\chi^{(2)}\alpha_1^*\alpha_2-i f_1\rho_{ge}+\epsilon_p e^{-i\omega t} , \label{eq:timea}
\\
\dot{\alpha}_2=(-i\omega_2-\gamma_2)\alpha_2-i\chi^{(2)}\alpha_1^2-i f_2\rho_{ge} , \label{eq:timeb}
\\
\dot{\rho}_{ge}=(-i\omega_{eg}-\gamma_{eg})\rho_{ge}+i(f_1\alpha_1+f_2\alpha_2)(\rho_{ee}-\rho_{gg}) , \label{eq:timec}
\\
\dot{\rho}_{ee}=-\gamma_{ee}\rho_{ee}+i \left[ f_1(\alpha_1^*+f_2\alpha_2^*)\rho_{ge}-(\alpha_1+\alpha_2)\rho_{ge}^*\right] , \label{eq:timed}
\end{align}
\end{subequations}
where $\gamma_1$, $\gamma_2$ are the damping rates of the MNP dimer modes $\alpha_1$, $\alpha_2$. $\gamma_{ee}$ and $\gamma_{eg}=\gamma_{ee}/2$ are the diagonal and off--diagonal decay rates of the quantum oscillator, respectively. To make a comparison, $\gamma_1$,$\gamma_2\sim 10^{14}$Hz for MNPs \cite{QDinducedtrans} while $\gamma_{ee}\sim 10^{12}$Hz for molecules \cite{spaser} and $\gamma_{ee}\sim 10^9$ Hz for quantum dots \cite{QCfano3}. The constraint on the conservation of probability $\rho_{ee}+\rho_{gg}=1$ accompanies Eqs.~(\ref{eq:timea}-\ref{eq:timed}).

%%%%%%%%%%%%%%%%%%%%%%%%%%%%%%%%%%%%%%%%%%%%%%%%%%%%%%%%%%%%%%%%%%%%%%%%%%%%%%%%%%%%%%%%%%%%%%%%%%%%%%%%%%%
%%%%%%%%%%%%%%%%%%%%%%%%%%%%%%%%%%%%%%%%%%%%%%%%%%%%%%%%%%%%%%%%%%%%%%%%%%%%%%%%%%%%%%%%%%%%%%%%%%%%%%%%%%%
%%%%%%%%%%%%%%%%%%%%%%%%%%%%%%%%%%%%%%%%%%%%%%%%%%%%%%%%%%%%%%%%%%%%%%%%%%%%%%%%%%%%%%%%%%%%%%%%%%%%%%%%%%%

%%%%%%%%%%%%%%%%%%%%%%%%%%%%%%%%%%%%%%%%%%%%%%%%%%%%%%%%%%%%%%%%%%%%%%%%%%%%%%%%%%%%%%%%%%%%%%%%%%%%%%%%%%%

In our simulations (Figs.~\ref{fig2}--\ref{fig4}), we time-evolve Eqs.~(\ref{eq:timea}-\ref{eq:timed}) numerically to obtain the long time behavior of $\rho_{eg}$, $\rho_{ee}$, $\alpha_1$, and $\alpha_2$. We determine the values to where they converge when the drive is on for long enough times. We perform this evolution for different $\omega_2$ frequency values with the initial conditions $\rho_{ee}(t=0)=0$, $\rho_{eg}(0)=0$, $\alpha_1(0)=0$, $\alpha_2(0)=0$. 

Beside the time-evolution simulations, one may gain understanding about the linear behavior of Eqs.~(\ref{eq:timea}-\ref{eq:timed}) by seeking solutions of the form
\begin{eqnarray}
&\alpha_1(t)=\tilde{\alpha}_1 e^{-i\omega t} \quad , \quad \alpha_2(t)=\tilde{\alpha}_2 e^{-i2\omega t} ,& \nonumber
\\
&\rho_{eg}(t)=\tilde{\rho}_{eg} e^{-i2\omega t} \quad , \quad \rho_{ee}(t)=\tilde{\rho}_{ee} ,& \label{eq:slowvary}
\end{eqnarray}   
for the steady states of the oscillations. This form of solutions are valid under the following assumptions. When the level spacing of the quantum oscillator is about the SH frequency, $\omega_{eg}\sim 2\omega$, its interaction with the first PP mode becomes highly off-resonant as compared to the $\hat{a}_2$ mode. In addition, MNP system can be chosen such that the hotspots of $\hat{a}_1$  mode and $\hat{a}_2$ mode emerge at different spatial positions. The quantum oscillator can be placed at the $\hat{a}_2$ hot-spot. In this case, its interaction with $\hat{a}_1$ mode can be neglected even without the need for the off-resonance assumption. In our numerical simulations governing the time-evolution of Eqs.~(\ref{eq:timea}-\ref{eq:timed}), we check that the solutions indeed converge to the form of Eq.~(\ref{eq:slowvary}) for long-time behavior. 

Inserting Eq.~(\ref{eq:slowvary}) into Eqs.~(\ref{eq:timea}-\ref{eq:timed}), one obtains the equations for the steady state
\begin{subequations}
\begin{align}
[i(\omega_1-\omega)+\gamma_1]\alpha_1+i2\chi^{(2)}\alpha_1^*\alpha_2=\epsilon_p , \label{eq:steadya}
\\
[i(\omega_2-2\omega)+\gamma_2]\alpha_2+i\chi^{(2)}\alpha_1^2=-i f_2\rho_{ge} , \label{eq:steadyb}
\\
[i(\omega_{eg}-2\omega)+\gamma_{eg}]\rho_{ge}=i f_2\alpha_2(\rho_{ee}-\rho_{gg}) , \label{eq:steadyc}
\\
\gamma_{ee}\rho_{ee}=i f_2(\alpha_2^*\rho_{ge}-\alpha_2\rho_{ge}^*) , \label{eq:steadyd}
\end{align}
\end{subequations}
where $\tilde{\alpha}_1$, $\tilde{\alpha}_2$, $\tilde{\rho}_{ge}$, $\tilde{\rho}_{ee}$ are constants independent of the time.

Using Eqs. (\ref{eq:steadyb}) and (\ref{eq:steadyc}), one can obtain the steady state value of $\hat{a}_2$ plasmon--polariton mode field as
\begin{equation}
\tilde{\alpha}_2=\frac{i\chi^{(2)}}{\frac{|f_2|^2 y}{i(\omega_{eg}-2\omega)+\gamma_{eg}}-[i(\omega_2-2\omega)+\gamma_2]}\;\tilde{\alpha}_1^2
\label{alph2}
\end{equation}
for $\omega_{eg}\sim 2\omega$. Here, $y=\rho_{ee}-\rho_{gg}$ is the steady state value of the population inversion. Quantum oscillator couples with the $\hat{a}_2$ mode into which SH conversion take place. Since the SH intensity is weak, quantum oscillator only weakly excited. Therefore, in Eq.~\ref{alph2} inversion usually takes on values very close to $y\simeq -1$. Even for enhanced SH conversion, as discussed in Sec.~\ref{sec:enhancement}, we observe in our simulations that the value of $y$ does not rise above -0.9. In our simulations, $y$ is not used as a fixed parameter with value $\sim-1$. We always determine $y$ from the time evolution of $\rho_{ee}-\rho_{gg}$.

%%%%%%%%%%%%%%%%%%%%%%%%%%%%%%%%%%%%%%%%%%%%%%%%%%%%%%%%%%%%%%%%%%%%%%%%%%%%%%%%%%%%%%%%%%%%%%%%%%%%%%%%%%%
\subsection{Suppression of the nonlinear conversion process} \label{sec:suppression}

Taking a closer look at the denominator of Eq.~(\ref{alph2}), one can immediately realize that $|f_2|^2 y/\gamma_{eg}$ attains huge values on resonance $\omega_{eg}=2\omega$. Because, linewidth of the quantum oscillator ($\gamma_{eg}$) is very small compared to all other frequencies. If $f_2\neq 0$, the largeness of the $|f_2|^2 y/\gamma_{eg}$ term dominates the denominator. This results in the suppression of the generation of the $\tilde{\alpha}_2$ plasmon-polariton polarization field in the MNP dimer. 

\begin{figure}
\includegraphics[width=3.2in]{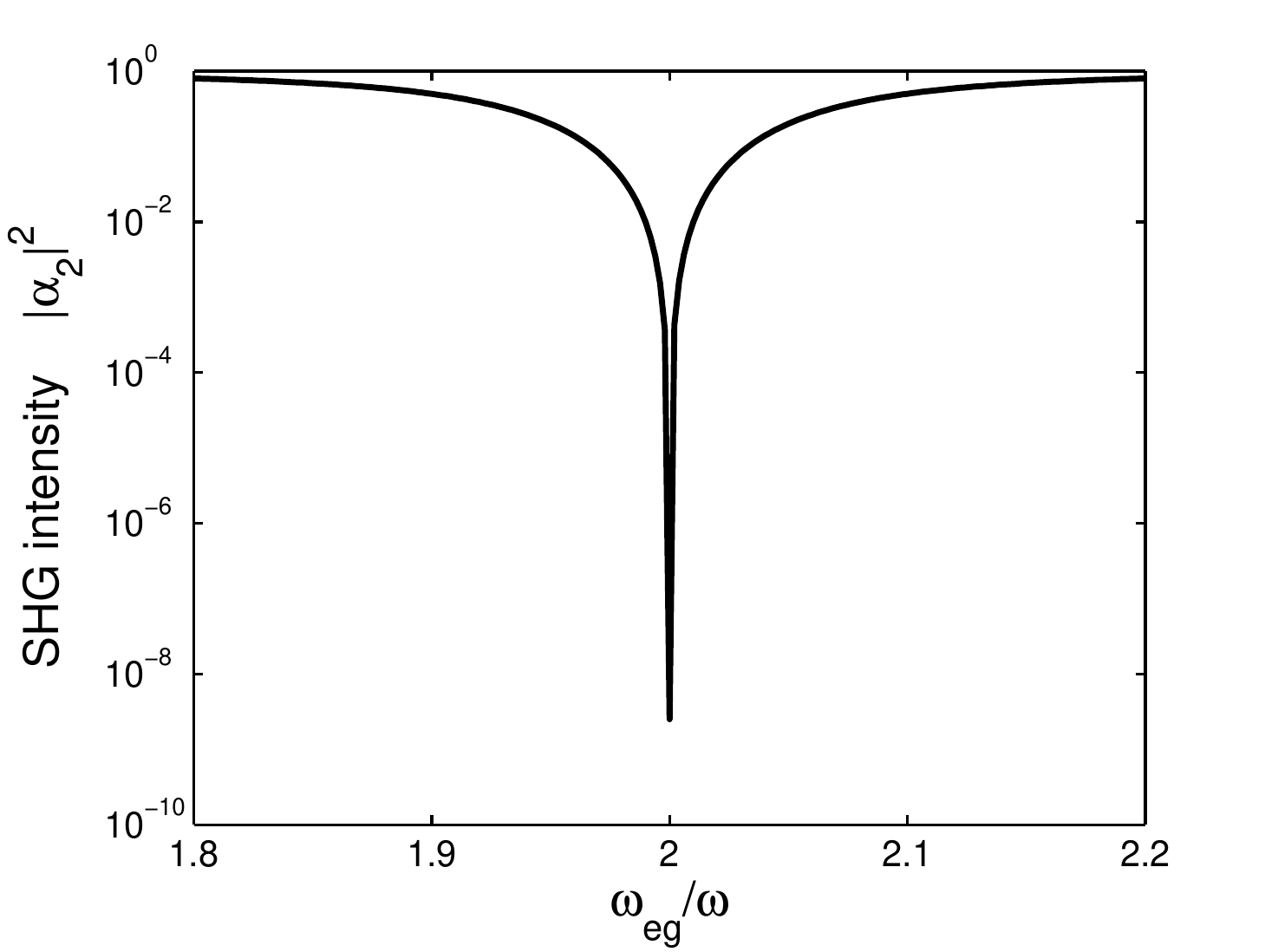}
\caption{Suppression of the SH polarization conversion to the $\hat{a}_2$ plasmon-polariton (PP) mode from the $\hat{a}_1$ mode. Even at the presence of resonant conversion condition, $\omega_1=\omega$ and $\omega_2=2\omega$, the presence of quantum oscillator ($\omega_{eg}=2\omega$) prevents the take place of the SHG process. EIT doesn't allow the polarization in the $\hat{a}_2$ PP mode. The resonant conversion is represented by unity in the figure. When $\omega_{eg}=2\omega$, the nonlinear intensity can be suppressed 9-orders of magnitude with respect to the resonant value. Decay rates are  $\gamma_1=\gamma_2=0.1\omega$ and $\gamma_{eg}=10^{-5}\omega$. We use $\chi^{(2)}=0.01\omega$ and $f_2=0.1\omega$.}
\label{fig2}
\end{figure}

In Fig.~\ref{fig2}, we demonstrate that the SHG in the MNP dimer can be suppressed very effectively by coupling the MNP dimer to a quantum oscillator. We time evolve Eqs.~(\ref{eq:timea}-\ref{eq:timed}) to obtain steady state values for the excitations.

\begin{figure}
\includegraphics[width=3.2in]{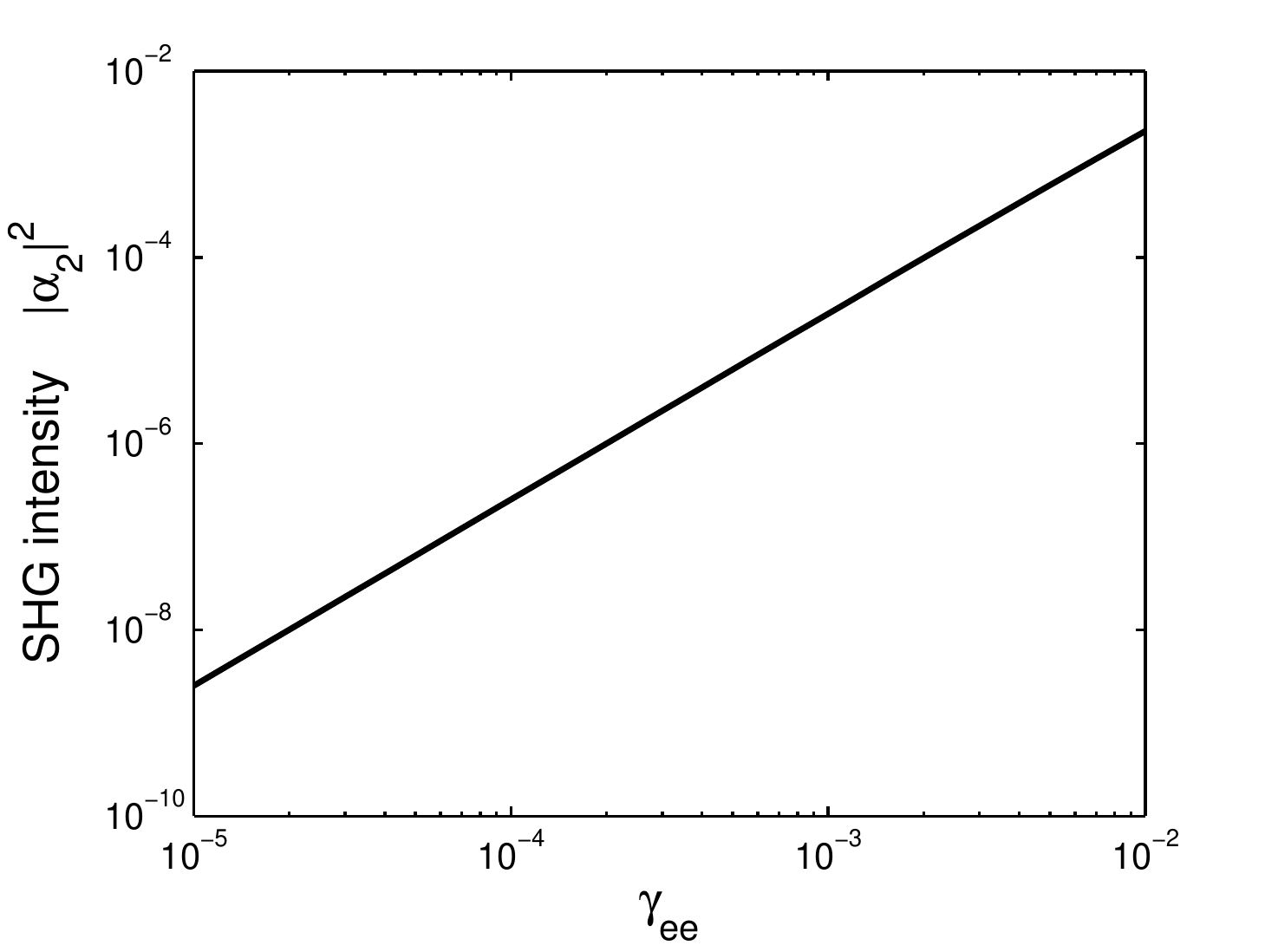}
\caption{The maximum suppression for different values of the quantum decay rate $\gamma_{ee}$. The log-log plot has a steepness of value $\simeq$2, thus pointing out the relation $|\alpha_2|^2\sim\gamma_{ee}^{2}$. This relation can be easily inferred from Eq.~(\ref{alph2}) for small values of  $\gamma_{ee}$ with $\omega_{eg}=2\omega$. Smaller quantum decay rate results in higher quality suppression.} 
\label{fig3}
\end{figure}

Without the presence of a quantum oscillator, the SHG would be maximum $(\tilde{\alpha}_2=-i\chi^{(2)}\tilde{\alpha}_1^2/\gamma_2)$ that is when the second plasmon-polariton mode is on resonance $\omega_2=2\omega$ [see Eq.~(\ref{alph2})]. In Fig.~\ref{fig2}, we observe that even at the presence of this resonance $(\omega_2=2\omega)$, EIT suppresses the SHG by 9 orders of magnitude. This effect arises simply because EIT doesn't allow the polarization of the second plasmon-polariton mode $\hat{a}_2$ at $2\omega$. The two paths --introduced in the MNP dimer due to the hybridization with the quantum oscillator-- for polarization transfer from the $\hat{a}_1$ mode interfere destructively. This cancels the transfer of $\hat{a}_1$ polarization (oscillating at $\omega$) to $\hat{a}_2$ polarization in the dimer (oscillating at 2$\omega$). Fig.~\ref{fig3} shows the dependence of the SHG intensity $|\alpha_2|^2$ on the decay rate $\gamma_{ee}$ of the quantum oscillator that is coupled to the dimer. The slope of the graph implies a $|\alpha_2|^2\sim\gamma_{ee}^2$ dependence. In fact, this can be easily inferred from Eq.~(\ref{alph2}) for the small values of $\gamma_{eg}$ when $\omega_{eg}\cong 2\omega$.

%
%

%%%%%%%%%%%%%%%%%%%%%%%%%%%%%%%%%%%%%%%%%%%%%%%%%%%%%%%%%%%%%%%%%%%%%%%%%%%%%%%%%%%%%%%%%%%%%%%%%%%%%%%%%%%
\subsection{Enhancement of the nonlinear conversion process} \label{sec:enhancement}

Contrary to suppression phenomenon, the interference effects can be arranged so that SHG process can be carried closer to the resonance. In the denominator of Eq.~(\ref{alph2}), the imaginary part of the first term $|f_2|^2 y/[i(\omega_{eg}-2\omega)+\gamma_{eg}]$ can be arranged to cancel the $i(\omega_2-2\omega)$ expression in the second term of the denominator. This gives the condition 
\begin{equation}
|f_2|^2 y(\omega_{eg}-2\omega)+(\omega_2-2\omega)[(\omega_{eg}-2\omega)^2+\gamma_{eg}^2]=0.
\label{eqcond}
\end{equation}
Eq.~(\ref{eqcond}) has two roots
\begin{equation}
\omega_{eg}^{(1,2)}-2\omega=\frac{|f_2|^2|y|}{\omega_2- 2\omega}\mp\sqrt{\frac{|f_2|^4|y|^2}{(\omega_2 -2\omega)^2}-4\gamma_{eg}^2}.
\label{eqroot}
\end{equation}
The first (smaller) root $\omega_{eg}^{(1)}\cong 2\omega+2\gamma_{eg}$ is not useful for SHG enhancement. Because  it enlarges the real part of the $|f_2|^2 y/[i(\omega_{eg}-2\omega)+\gamma_{eg}]$ term. This is already the suppression condition for SHG.

Since $\omega_{eg}$ is not very close to $2\omega$ for the second root $\omega_{eg}^{(2)}$, it does not cause the real part of the $|f_2|^2 y/[i(\omega_{eg}-2\omega)+\gamma_{eg}]$ term to rapidly diverge. At the same time, $\omega_{eg}^{(2)}$ is minimizes the absolute value of the denominator of Eq.~(\ref{alph2}), that gives rise to the maximum SHG. 

For the case of the suppression of SHG, one can safely use the approximation $y\cong-1$, because excitations are suppressed in the hybrid system, $\rho_{ee}\cong0$, and this leads to $y=\rho_{ee}-\rho_{gg}\cong-1$. However, in the case of SHG enhancement, one can not use the value $\cong-1$ for $y$. We observe that at resonances $y$ can attain inversion values that are close to zero. Nevertheless, Eq.~(\ref{eqroot}) still serves at least as a guess value for the order of  $\omega_{eg}^{(2)}$, where SHG enhancement arises. 

\begin{figure}%[H]
\includegraphics[width=3.2in]{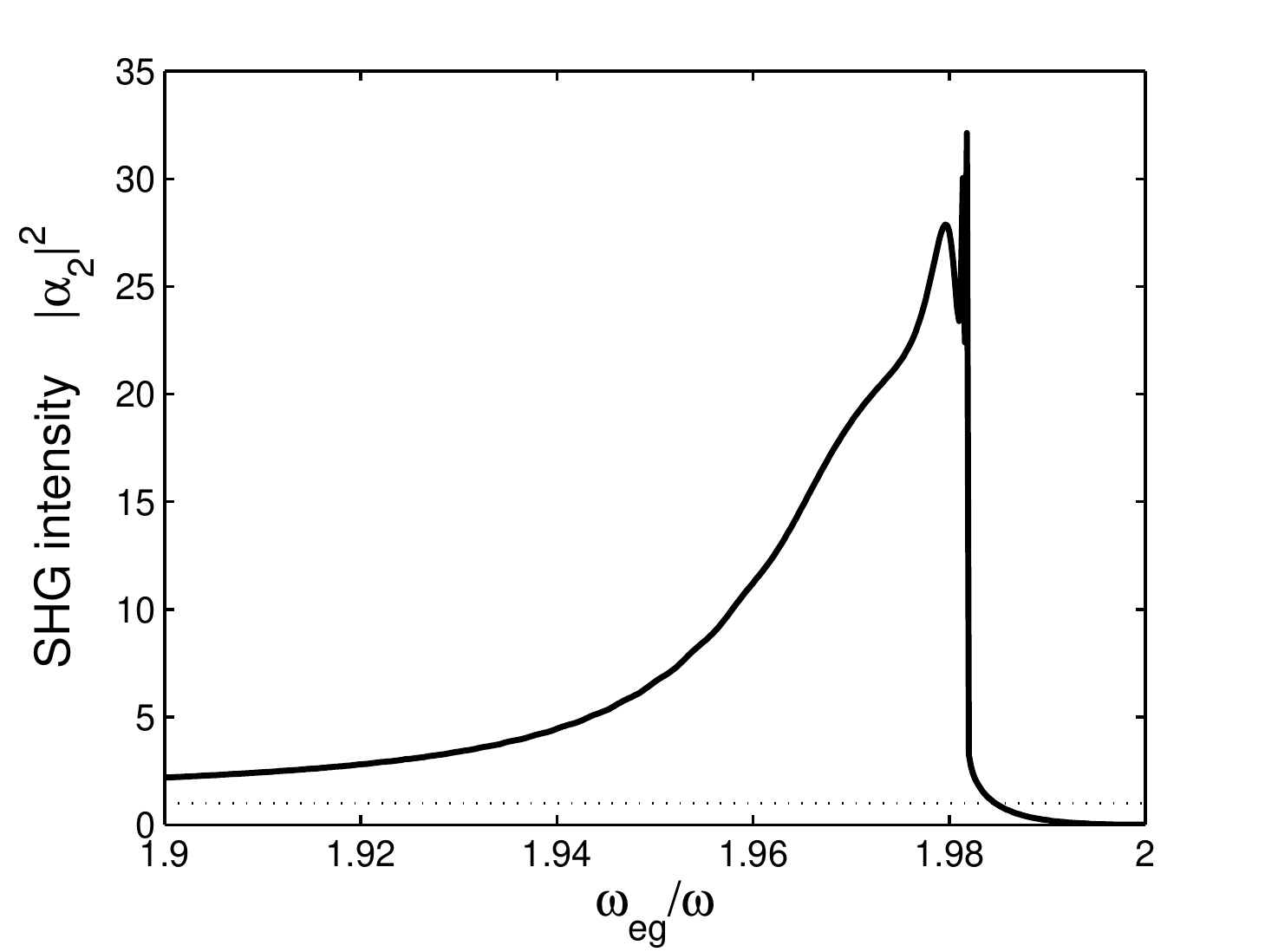}
\caption{The enhancement of the nonlinear process. The second PP mode ($\hat{a}_2$) is far-off resonant to the SHG ($\omega_2=2.4\omega$). The nonlinear process can be carried closer to resonance by arranging the quantum level spacing \cite{PSarrangeweg} to $\omega_{eg}\cong1.98\omega$. The conversion is enhanced upto 30 times compared to the off-resonant process. The conversion for off-resonant process (with $f_2=0$) is represented by unity in the figure. For $\omega_{eg}=2\omega$, nonlinear process is suppressed similar to Fig.~\ref{fig2}. Note that, the bandwidth for the enhancement of the nonlinear conversion is much wider than the suppression bandwidth. Decay rate of the nanoscale resonator \cite{hqcavity1,hqcavity2} is chosen as $\gamma_2=0.01\omega$. We use $\chi^{(2)}=0.01\omega$ and $f_2=0.1\omega$.}
\label{fig4}
\end{figure}

In Fig~\ref{fig4}, we depict the enhancement of the nonlinear frequency conversion for a nanoscale dimer whose PP mode ($\omega_2=2.4\omega$) is far off-resonant to the SH frequency. Off-resonant SHG conversion [$f_2=0$ in Eq.~(\ref{alph2})] is represented by unity in Fig.~\ref{fig4}. We observe a 30 times enhancement in the conversion intensity for the choice of $\omega_{eg}=1.98\omega$. Parameters are given in Fig.~\ref{fig4}. As it can be inferred from Eq.~(\ref{alph2}); relative enhancement efficiency can be grown much more with respect to the off-resonant value ($\omega_2=2.4\omega$ with $f_2=0$) if higher quality (small $\gamma_2$) resonators are used. Quality factors of $\sim$1300 can be achieved for micro-cavities operating at optical wavelength \cite{hqcavity1,hqcavity2}.

In obtaining Eqs.~(\ref{eq:steadya})-(\ref{eq:steadyd}), we neglect the coupling between the $\hat{a}_1$ PP mode with the quantum oscillator due to off-resonant behavior. Without such a negligence, analytic results like Eqs.~(\ref{eq:steadya})-(\ref{eq:steadyd}) cannot be obtained since expressions in Eq.~(\ref{eq:slowvary}) are not valid anymore. In case, when quantum level spacing $\omega_{eg}$ is close to $\omega_1$, steady state value of $\alpha_2$ is obtained numerically by time evolution of Eqs.~(\ref{eq:timea})-(\ref{eq:timed}). Surprisingly, enhancement factor obtained in this case can reach as high as $\sim$1000. Such an enhancement factor, consistently, shown of being able to explain the observed SH signal, from noncentrosymmetric MNP clusters decorated with molecules by CW laser irradiation \cite{BekSHG}.

For the suppression effect to emerge, the first term in the denominator of Eq.~(\ref{alph2}) must be sufficiently large enough. This necessitates the coupling of plasmonic resonator to a quantum oscillator which has a small decay rate $\gamma_{eg}$. On the other hand, cancellation of the nonresonant terms in Eq.~(\ref{eqcond}) does not require the presence of a high-quality oscillator with a sharp resonance. Hence, nonlinear response enhancement may emerge even for coupled plasmonic resonators with broad emission bands.

%%%%%%%%%%%%%%%%%%%%%%%%%%%%%%%%%%%%%%%%%%%%%%%%%%%%%%%%%%%%%%%%%%%%%%%%%%%%%%%%%%%%%%%%%%%%%%%%%%%%%%%%%%%
\subsection{Comparison with 3D simulations} \label{sec:MNPBEM}

\begin{figure*}
\includegraphics[width=6.4in]{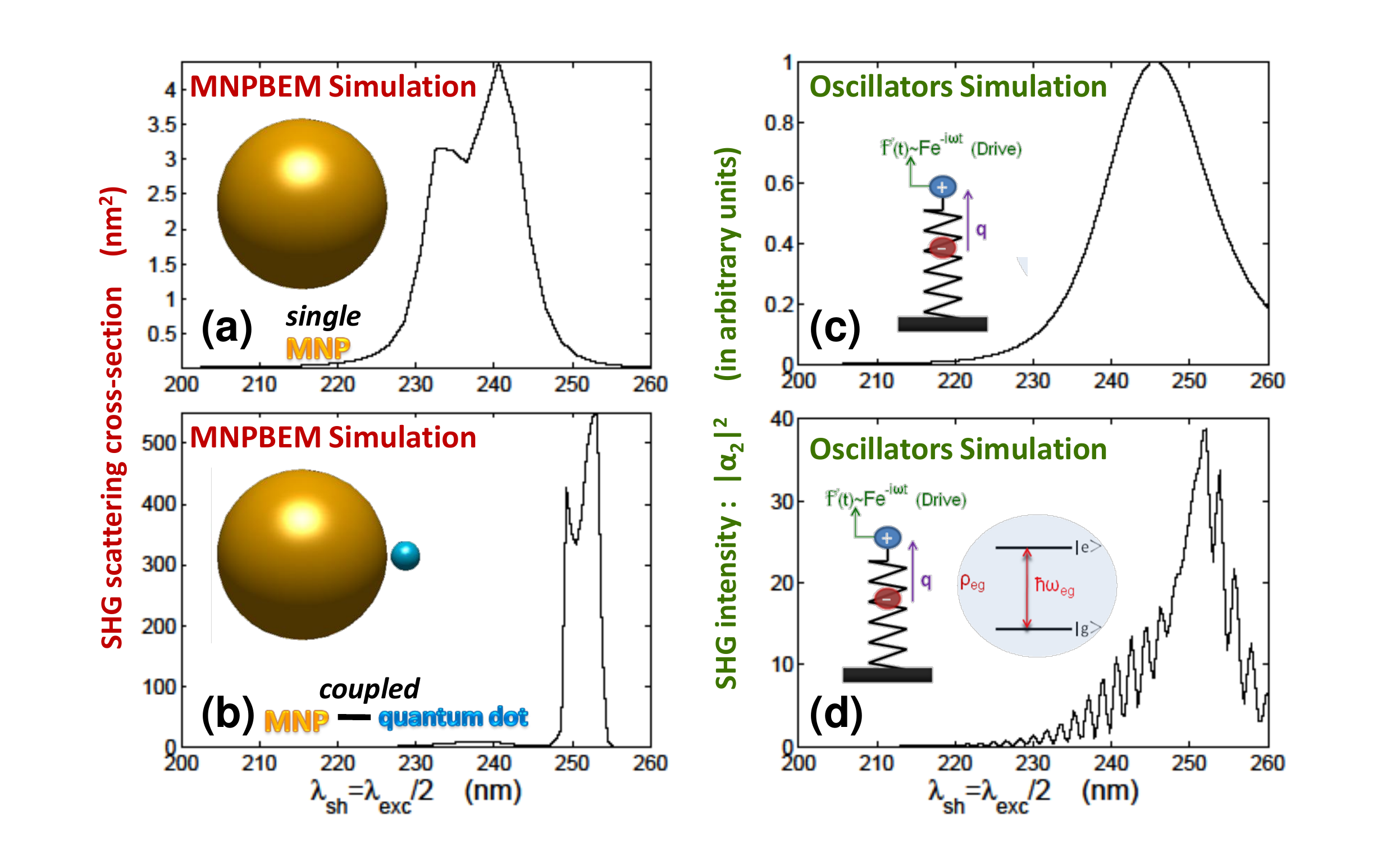}
\caption{(a,b) 3D MNPBEM calculation of second harmonic (SH) scattering cross-section (a) when a gold nanoparticle stands alone and (b) when the gold nanoparticle interacts with a small object (blue) which has a sharp resonance ($\gamma_{eg}\ll \gamma_1$) at $\lambda_{eg}=$500nm. Due to the Fano resonance at $\lambda_{\text{exc}}\gtrsim\lambda_{eg}=$500nm ($\lambda_{\text{sh}}\gtrsim\lambda_{eg}/2=$250nm)  SHG is enhanced about 100 times in the presence of the blue particle with small decay rate; even though its linear extinction cross-section does not increase. (c,d) Simulation of coupled classical (low quality) and quantum (high quality) oscillators using Eq.s (\ref{eq:timea}-\ref{eq:timed}). Intensity of the SH field (c) when a classical oscillator is driven alone and (d) when an quantum oscillator which has a tiny decay rate interacts with the driven oscillator. The simple model of coupled oscillators predicts the take place of SHG enhancement and the spectral position of the Fano resonance correctly. Diameters of gold nanoparticle and quantum oscillator (a QD) are 70nm and 12nm, respectively. The gap size is 1.5nm. Other parameters used in the simulation are $\lambda_1=490$nm, $\lambda_2=250$nm (plasmon extinction peaks of the single MNP), $f_1=0.03\omega_1=0.03\times2\pi c/\lambda_1$, $\chi^{(2)}=0.01\omega_1$.   }
\label{fig5}
\end{figure*}

In Fig.~\ref{fig5}, we compare the predictions of our model [Eq.s (\ref{eq:timea}-\ref{eq:timed})] with the 3-dimensional boundary element simulations. Simulations are performed with the MNPBEM toolbox \cite{MNPBEM} in Matlab. We use the {\it bemret}  environment \cite{MNPBEM} which is based on the exact solutions of the Maxwell equations using surface integral evaluations \cite{AbajoPRB2002}. Simulations take the retardation effects into account. We use the experimental data, that is present in the toolbox, for the dielectric function of the gold nanoparticles. Coupled oscillators model correctly predicts the emergence of the SHG enhancement (compare Fig.~\ref{fig5}b and \ref{fig5}d) as well as its position in the spectrum. 

%%
%%
%\begin{figure*}
%\includegraphics[width=6.4in]{fig5}
%\caption{(a,b) 3D MNPBEM calculation of second harmonic (SH) scattering cross-section (a) when a gold nanoparticle stands alone and (b) when the gold nanoparticle interacts with a small object (blue) which has a sharp resonance ($\gamma_{eg}\ll \gamma_1$) at $\lambda_{eg}=$500nm. Due to the Fano resonance at $\lambda_{\text{exc}}\gtrsim\lambda_{eg}=$500nm ($\lambda_{\text{sh}}\gtrsim\lambda_{eg}/2=$250nm)  SHG is enhanced about 100 times in the presence of the blue particle with small decay rate. (c,d) Simulation of coupled classical (low quality) and quantum (high quality) oscillators using Eq.s (\ref{eq:timea}-\ref{eq:timed}). Intensity of the SH field (c) when a classical oscillator is driven alone and (d) when an quantum oscillator which has a tiny decay rate interacts with the driven oscillator. The simple model of coupled oscillators predicts the take place of SHG enhancement and the spectral position of the Fano resonance correctly.}
%\label{fig5}
%\end{figure*}
%%
%%

In the MNPBEM simulation, we calculate the SH response of a gold nanoparticle of 70nm diameter. Fig.~\ref{fig5}a shows the emitted SH radiation in case the gold nanoparticle stands alone. The driving radiation $\lambda_{\text{exc}}=2\lambda$ is converted to the SH wavelength $\lambda$. In Fig.~\ref{fig5}b, we place a small particle (blue), which has a tiny decay rate $\gamma_{eg}\ll \gamma_1$, near the gold nanoparticle. The small blue sphere, of 12nm diameter, is filled with a dielectric medium which has a Lorentzian response function $\epsilon(\omega)=1+\omega_{p}^2/(\omega_{eg}^2-\omega^2-i\gamma_{eg}\omega)$, where $\omega_p$ determines the oscillation (polarization) strength. In this way, the small blue particle mimics the response of a dye molecule having a sharp resonance ($\gamma_{eg}\ll\gamma_1$) near $\omega_{eg}$ ($\lambda_{eg}=$500nm). Fig.~\ref{fig5}b shows that SHG is enhanced about 100 times near $\lambda_{eg}/2$ for $\lambda_{\text{exc}}\gtrsim \lambda_{eg}=$500nm. The take place of such a phenomenon for $\lambda_{sh}^*=\lambda_{\text{exc}}/2  \simeq $255nm can be predicted within the simulations of our model. The SHG intensity for the classical oscillator (Fig.~\ref{fig5}c) is enhanced about 40 times at the same spectral position (Fig.~\ref{fig5}d). A careful glance at Fig.~\ref{fig5}b reveals that the normal SHG remains at $\lambda=$240nm. 

It is also worth noting that, a linear Fano resonance \cite{QCfano1,metasginNanoscale2013} reveals itself in the form of a dip at the center of the scattering cross-section peak (at $\lambda_{eg}=500$nm) in BEM simulation. The fact that position of the dip follows the resonance $\lambda=\lambda_{eg}=2\pi c/\omega_{eg}$, ensures that this effect is indeed a Fano resonance --not a simple hybridization between the two particles.

In Fig.~\ref{fig5}, we demonstrate the enhancement of SHG for $\omega_{eg}\sim \omega_1\sim\omega$, in difference to the case we discuss in Eq. (\ref{alph2}) and Fig.~\ref{fig3} where $\omega_{eg}\sim 2\omega\sim\omega_2$. MNPBEM toolbox is limited in handling Fano resonances for $\omega_{eg}\sim 2\omega$. It treats the second power of the electric field's normal component at the(inner) particle boundaries as SH source and calculates the SH field intensity \cite{MNPBEM}. Since SH field ($2\omega$) is weak compared to the linear ($\omega$) response and $\omega_{eg}\sim \omega$ is off-resonant to $2\omega$, Fano resonances near $\omega_{eg}\sim \omega$ can be treated within this approach. On the contrary, Fano resonances near $\omega_{eg}\sim 2\omega$ cannot be handled within this method. Because, presence of a particle (blue) resonant to $\sim 2\omega$ cannot alter (has no feedback on) the magnitude of the generated SH sources which are only the second power of the linear ($\omega$) electric field distribution.

Nevertheless, Fig.~\ref{fig5} reveals the reliability of our model which can explain the underlying physical phenomenon leading to control over the SH response. The Fano resonance effects for $\omega_{eg}\sim 2\omega$, depicted in Fig.s~\ref{fig2} and \ref{fig3}, would be observed for a more complete treatment studied in Ref. \cite{KauranenOptExpress2011}.

Here, we discuss a single MNP coupled to a high quality oscillator, in difference to 2 MNPs configuration presented in Fig.~\ref{fig1}. The arrangement pictured in Fig.~\ref{fig1} provides stronger coupling between MNPs and quantum oscillator. However, 2 MNPs give two scattering resonances about $\omega_1$, which complicates the comparison between MNPBEM simulation and simple oscillators simulation. The double peak in Fig.~\ref{fig5}a is due to the emergence of double resonance near $\omega_2$ even for a single MNP. The coupled system is excited with an x-polarized (electric field is along the line connecting the centers of the MNP and quantum dot) plane wave. When y- or z-polarized plane wave excitation is used, no enhancement of SH conversion is obtained. 

\begin{figure}
\includegraphics[width=3.4in]{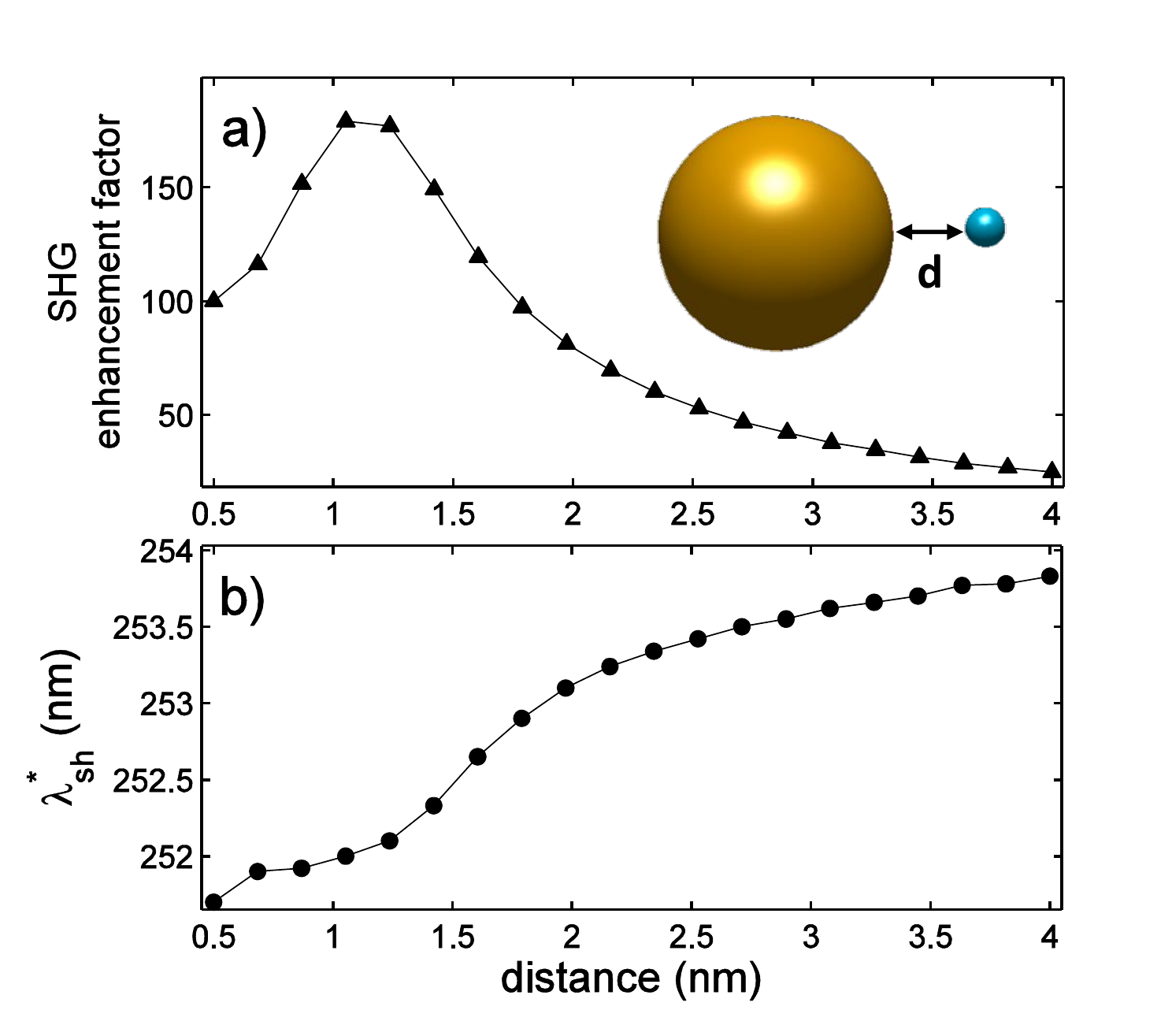}
\caption{ MNPBEM simulation. (a) Dependence of the enhancement factor for second harmonic (SH) conversion to the gap size, $d$, and (b) corresponding excitation wavelengths $\lambda_{ex}=2\lambda_{sh}^{*}$ where enhancement emerges. The system is the same with Fig.~\ref{fig5}. Only the distance between MNP and quantum oscillator (a QD) is varied. }
\label{fig6}
\end{figure}

In Fig.~\ref{fig6}a, we show the dependence of the SHG enhancement factor to the size of the gap between the MNP and the quantum dot (QD). In Fig.~\ref{fig6}b, variation of the corresponding maximum enhancement wavelength as a function of gap size is given.

%
%

%%%%%%%%%%%%%%%%%%%%%%%%%%%%%%%%%%%%%%%%%%%%%%%%%%%%%%%%%%%%%%%%%%%%%%%%%%%%%%%%%%%%%%%%%%%%%%%%%%%%%%%%%%%
\subsection{Model for other nonlinearities} \label{sec:otherNL}

In case the control of the third harmonic generation (THG) is required, instead of SHG, change of Eq.~(\ref{eq:Hsh}) to
\begin{equation}
\hat{H}_{th}=\hbar\chi^{(3)} \left( \hat{a}_2^{\dagger} \hat{a}_1 \hat{a}_1\hat{a}_1 + \hat{a}_1^{\dagger} \hat{a}_1^{\dagger}\hat{a}_1^{\dagger} \hat{a}_2\right),
\end{equation} 
would be sufficient. Eq.~(\ref{alph2}) should simply be modified to
\begin{equation}
\tilde{\alpha}_2=\frac{i\chi^{(3)}}{\frac{|f_2|^2 y}{i(\omega_{eg}-3\omega)+\gamma_{eg}}-[i(\omega_2-3\omega)+\gamma_2]}\;\tilde{\alpha}_1^3.
\label{eq:alpha2TH}
\end{equation}
All of our considerations, stated above, works for Eq.~(\ref{eq:alpha2TH}) as well. In order to be able to use the introduced model, one has to be careful if the nonlinear conversion oscillates another PP mode which is resolvable from the first one.

In some nonlinear conversion processes (e.g. enhanced Raman scattering \cite{Sharma2012}), both the drive ($\omega$) and the generated ($\omega_{NL}$) frequencies may excite the same PP mode ($\hat{a}_1$). This may occur because, frequency spacing $|\omega_{NL}-\omega|$ can be small compared to the decay rate ($\gamma_1$) of the PP polarization mode. In this case too, suppression of the nonlinear conversion will necessarily arise for $\omega_{eg}=\omega_{NL}$. This is because; coupling to a high quality factor oscillator prevents the polarization/absorption for frequency values around $\simeq\omega_{eg}$ independent of the details of the conversion mechanism \cite{metasginNanoscale2013,Alzar,Soukoulis2012}. The generated frequency mode will be suppressed if it coincides with $\omega_{eg}$. Fig. 5 in ref. \cite{metasginNanoscale2013} demonstrates that polarization cancellation emerges at $\omega=\omega_{eg}$ for the coupled MNP-quantum oscillator system. However, enhancement of the nonlinearities depends on the physics of the conversion mechanism.

%%%%%%%%%%%%%%%%%%%%%%%%%%%%%%%%%%%%%%%%%%%%%%%%%%%%%%%%%%%%%%%%%%%%%%%%%%%%%%%%%%%%%%%%%%%%%%%%%%%%%%%%%%%
%%%%%%%%%%%%%%%%%%%%%%%%%%%%%%%%%%%%%%%%%%%%%%%%%%%%%%%%%%%%%%%%%%%%%%%%%%%%%%%%%%%%%%%%%%%%%%%%%%%%%%%%%%%
%%%%%%%%%%%%%%%%%%%%%%%%%%%%%%%%%%%%%%%%%%%%%%%%%%%%%%%%%%%%%%%%%%%%%%%%%%%%%%%%%%%%%%%%%%%%%%%%%%%%%%%%%%%

\section{Classification of Fano Resonances} \label{sec:classification}

 Fano resonances induced in coupled classical/quantum oscillators follow from a common mechanism. Coupling of the resonant excitation to an auxiliary mode (QD in our case) introduces two possible absorption paths. The two paths counteract and avoid the excitation (hence polarization and absorption) at the resonance \cite{Alzar}. The same mechanism is also responsible for the phenomenon of electromagnetically induced transparency (EIT) \cite{Scullybook} in 3-level atoms. Path interference effect can reveal itself in a variety of systems. Therefore, a classification of such coupled systems becomes necessary, for the purpose of illuminating the possible extensions of the nonlinearity enhancement to other systems.

The first class can be defined as follows. A classical/quantum oscillator can be coupled to a classical or quantum object. Here, the word {\it coupling} implies an interaction which does not yield a strong hybridization that can change the entire spectrum severely. In other words, the splitting in the plasmon mode (which has a wide spectral width) is below the resolution limit. This class contains the following examples. (i) A MNP attached with a molecule/QD, (ii) a molecule exhibiting nonlinear response attached to a QD \cite{molecule_QD}, or (iii) two weakly interacting MNPs (e.g. distance between two MNPs is relatively long or a dielectric is placed in between).
     
     The second class is closer, in physical aspects, to the standard EIT. Coupling is strong enough to completely modify the spectrum, e.g. two closely placed hybridized MNPs (dimer). In this case, too, Fano resonance can be induced if the bright (dipole-like) and dark (higher orders) modes overlap spectrally \cite{hybridFano1,hybridFano2} and spatially --i.e. overlap integral for the interaction Hamiltonian does not vanish (similar to Ref.~\cite{overlapintegral}). This is a single system whose internal states interact. In this respect, this class is analog of EIT. The difference is; in EIT, the third level (dipole-forbidden transition) --first level is the ground state, second is the dipole-allowed excited state-- is coupled externally with the excited level using a microwave drive \cite{Scullybook}. In the plasmonic analog, coupling between the two excitations arises naturally due to the spectral overlap, which is absent in 3-level atoms.
     
%Consequently, if very strong coupling between two 2-level quantum objects would be achieved, an external drive (e.g. microwave as in 3-level atoms) would be necessary. Because, the hybridized states would be sharp and would not interact with each other unlike Ref. [Remo]. Thus, the absorption dip at the resonance for this system would be called as EIT, instead of Fano resonance.

%%%%%%%%%%%%%%%%%%%%%%%%%%%%%%%%%%%%%%%%%%%%%%%%%%%%%%%%%%%%%%%%%%%%%%%%%%%%%%%%%%%%%%%%%%%%%%%%%%%%%%%%%%%
%%%%%%%%%%%%%%%%%%%%%%%%%%%%%%%%%%%%%%%%%%%%%%%%%%%%%%%%%%%%%%%%%%%%%%%%%%%%%%%%%%%%%%%%%%%%%%%%%%%%%%%%%%%
%%%%%%%%%%%%%%%%%%%%%%%%%%%%%%%%%%%%%%%%%%%%%%%%%%%%%%%%%%%%%%%%%%%%%%%%%%%%%%%%%%%%%%%%%%%%%%%%%%%%%%%%%%%

\section{Discussions and Conclusions} \label{sec:conclusion}

It is well demonstrated that the presence of a quantum oscillator, with a smaller decay rate, changes the optical response of MNPs dramatically. Due to the destructive interference of the (hybridized) absorption paths, MNP can not be polarized at the resonance frequency of the quantum oscillator. 

We demonstrate that a similar path interference effect can be adopted to both suppress and enhance the nonlinear conversion processes in a MNP dimer. A quantum oscillator is placed in the center of two hybridized MNPs where hot-spot of both plasmon-polariton mode emerges. If the quantum oscillator is resonant to the second or third harmonic frequency, (e.g. $\omega_{eg}=2\omega$), this frequency conversion process is suppressed several orders of magnitude. Because, EIT does not allow the excitation of the $2\omega$ oscillation in the second plasmon-polariton mode of the dimer ($\omega_2$). On the other hand, the similar interference effects can be used also to enhance the nonlinear frequency conversion. The level spacing of the quantum oscillator can be arranged \cite{PSarrangeweg} so that the nonresonant [e.g. ($\omega_2-2\omega$)] terms cancel. 

It is worth noting that emergence of the enhancement phenomenon does not require coupling to a high-quality oscillator. A small decay rate ($\gamma_{eg}$) is not necessary for nonresonant terms to cancel in Eq.~(\ref{eqcond}). Hence, the effect can readily be observed also for two coupled low-quality MNPs, if one chooses the resonances of the nanoparticles properly.

We compare the predictions of our simple model with 3D MNPBEM simulations which are based on the exact (computational) calculations of the Maxwell equations. We show that our model successfully predicts the emergence of the SHG enhancement as well as its position in the spectrum.

We present our method for the engineering of SHG process. However, the method can be used also for other nonlinear frequency generation processes. This happens as long as the converted frequency falls in the range of a different plasmon-polariton mode. Contrary to the enhancement case, suppression phenomenon is independent of the conversion mechanism. If a quantum oscillator --resonant to the generated frequency-- is coupled to the MNP system, the conversion process is prohibited. Hence, it can be used to suppress undesired Raman scattering processes which cause loses in the signal strength.

%%%%%%%%%%%%%%%%%%%%%%%%%%%%%%%%%%%%%%%%%%%%%%%%%%%%%%%%%%%%%%%%%%%%%%%%%%%%%%%%%%%%%%%%%%%%%%%%%%%%%%%%%%%
%%%%%%%%%%%%%%%%%%%%%%%%%%%%%%%%%%%%%%%%%%%%%%%%%%%%%%%%%%%%%%%%%%%%%%%%%%%%%%%%%%%%%%%%%%%%%%%%%%%%%%%%%%%
%%%%%%%%%%%%%%%%%%%%%%%%%%%%%%%%%%%%%%%%%%%%%%%%%%%%%%%%%%%%%%%%%%%%%%%%%%%%%%%%%%%%%%%%%%%%%%%%%%%%%%%%%%%
%%%%%%%%%%%%%%%%%%%%%%%%%%%%%%%%%%%%%%%%%%%%%%%%%%%%%%%%%%%%%%%%%%%%%%%%%%%%%%%%%%%%%%%%%%%%%%%%%%%%%%%%%%%
%%%%%%%%%%%%%%%%%%%%%%%%%%%%%%%%%%%%%%%%%%%%%%%%%%%%%%%%%%%%%%%%%%%%%%%%%%%%%%%%%%%%%%%%%%%%%%%%%%%%%%%%%%%
%%%%%%%%%%%%%%%%%%%%%%%%%%%%%%%%%%%%%%%%%%%%%%%%%%%%%%%%%%%%%%%%%%%%%%%%%%%%%%%%%%%%%%%%%%%%%%%%%%%%%%%%%%%

\begin{acknowledgements}
M.E.T and D.T. acknowledge  support  from T\"{U}B\.{I}TAK-KAR\.{I}YER  Grant No.  112T927. This work was undertaken while one of the authors (M.E.T) was in residence at Bilkent University with the support provided by  O\u{g}uz G\"{u}lseren. A.B. acknowledges support from Bilim Akademisi  The Science Academy, Turkey under the BAGEP program and METU BAP-08-11-2011-129 grant. The research leading to these results has received funding from the European Union's Seventh Framework Program FP7/2007-2013 under grant agreement no. 270483.
\end{acknowledgements}

%%%%%%%%%%%%%%%%%%%%%%%%%%%%%%%%%%%%%%%%%%%%%%%%%%%%%%%%%%%%%%%%%%%

%%%%%%%%%%%%%%%%%%%%%%%%%%%%%%%%%%%%%%%%%%%%%%%%%%%%%%%%%%%%%%%%%%
%%%%%%%%%%%%%%%%%%%%%%%%%%%%%%%%%%%%%%%%%%%%%%%%%%%%%%%%%%%%%%%%%%
%%%%%%%%%%%%%%%%%%%%%%%%%%%%%%%%%%%%%%%%%%%%%%%%%%%%%%%%%%%%%%%%%%%
%\begin{appendix}
%\section{Equations of Motion}
%
%
%
%\end{appendix}
%%%%%%%%%%%%%%%%%%%%%%%%%%%%%%%%%%%%%%%%%%%%%%%%%%%%%%%%%%%%%%%%%%
%%%%%%%%%%%%%%%%%%%%%%%%%%%%%%%%%%%%%%%%%%%%%%%%%%%%%%%%%%%%%%%%%%

\begin{thebibliography}{99}
\bibitem{PS5} SH conversion of molecules cannot be emerging due to the field enhancement effect of the MNPs, as well. This is because, in Ref.~\cite{BekSHG} we note that no SH signal is detected from centrosymmetric MNP clusters decorated with molecules.
\bibitem{PShighQ} Use of resonators with high quality behavior (Q=1300 in ref~\cite{hqcavity1,hqcavity2}) can provide narrower bands for suppression/enhancement of frequency conversion.
\bibitem{PSarrangeweg} In practice, arrangement of the level spacing of a quantum oscillator ($\omega_{eg}$) to high accuracy may not be possible for experimental demonstrations. However one may arrange the incident frequency $\omega$ such that $2\omega$ coincides with the desired level spacing.
\end{thebibliography}
\end{document}